\documentclass[twocolumn,showpacs,prl,10pt]{revtex4} 
\usepackage[english]{babel}
\usepackage{amsmath,amssymb}
\usepackage{times}
\usepackage{epsfig} 

\begin{document} 
\title{Fermi surface and pseudogap in electron-doped cuprates}

\author{M. M. Zemlji\v c$^{1}$, P. Prelov\v sek$^{1,2}$ and
T. Tohyama$^{3}$}
\affiliation{$^1$J.\ Stefan Institute, SI-1000 Ljubljana, Slovenia}
\affiliation{$^2$ Faculty of Mathematics and Physics, University of
Ljubljana, SI-1000 Ljubljana, Slovenia}
\affiliation{$^3$ Yukawa Institute for Theoretical Physics, Kyoto University,
Kyoto 606-8502, Japan}

\date{\today}

\begin{abstract}
Spectral functions are evaluated numerically within the $t$-$t'$-$J$
model as relevant for electron-doped cuprates. The Fermi surface
develops from a pocket-like into a large one with doping. The
corresponding pseudogap in the nodal direction is well resolved at
low doping and vanishes at intermediate doping. Its presence is strongly
temperature dependent and thus connected to longer-range
antiferromagnetic correlations. Features including double effective
band and optical conductivity are consistent with experiments on
electron-doped cuprates without invoking the closing of the
Mott-Hubbard gap.
\end{abstract} 

\maketitle 

The insight into the interesting and challenging physics of
high-temperature superconductors (HTSC) has been greatly enhanced by
angle-resolved photoemission spectroscopy (ARPES) experiments on
cuprates \cite{dama}. One of the challenging aspects is a pronounced
asymmetry of spectral properties between more investigated hole-doped
and electron-doped cuprates (EDC), as revealed by recent ARPES experiments
on representatives of the latter materials, as Nd$_{2-x}$Ce$_x$CuO$_4$
(NCCO) \cite{armi2,mats} and Sm$_{2-x}$Ce$_x$CuO$_4$ (SCCO)
\cite{park}. While the n-type parent compound Nd$_2$CuO$_4$ has very
similar excitation spectra to p-type undoped cuprates, the development
with doping is different. At low electron doping $x\sim 0.04$ the
ARPES Fermi surface (FS) reveals pronounced pockets at ${\bf
k}=(\pi,0)$ consistent with doubling of the unit cell due to the
long-range antiferromagnetic (AFM) order \cite{armi2}. With increasing
$x$ additional dispersive band-like excitation appears at $\omega<0$
and reaches the chemical potential ($\omega=0$) along the diagonal ${\bf k} \sim
(\pi/2,\pi,2)$ at intermediate doping $x=0.15$, which coincides with
the disappearance of long-range AFM order and the onset of
superconductivity. Consequently, the FS transforms into a large one although
still with well resolved quasiparticles only along parts of the FS
\cite{armi2,mats,park}. More recent analyses of NCCO \cite{mats} and SCCO
\cite{park} at intermediate doping seem to resolve both effective
bands and found the connection with the pseudogap observed in optical
conductivity \cite{onos,park}. While in the latter the temperature
dependence of the pseudogap reveals relation to the 
AFM correlations, there is so far no temperature-dependent ARPES data on EDC which
would directly relate the ARPES pseudogap to AFM ordering.

On the theoretical side, the EDC represent another very stringent test
for microscopic models of HTSC materials. It has become increasingly
clear that in prototype Hubbard $t$-$U$ model and reduced $t$-$J$
model additional second-neighbor hopping $t'$ (as well as
third-neighbor $t''$) can account well for the asymmetry in several
quantities \cite{tohy1,tohy2}. Still, the origin of the FS transformation from a
pocket-like to a large one as found by ARPES remains the central open
issue. First theories were based on the strongly renormalized interaction $U$
\cite{kusu} and the scenario of
doping-decreasing $U$ was proposed which results in both Hubbard bands
crossing the FS at intermediate doping \cite{kusk,sene}. More recent
slave-boson approach \cite{yuan} and numerical calculations within the
Hubbard model, using the variation-cluster \cite{aich} and
dynamical-cluster approaches \cite{macr} give indication that
most features of FS reconstruction can be reproduced with
doping-independent $U$. More detailed insight into the spectral
properties has been obtained from the numerical exact-diagonalization (ED)
study of $t$-$t'$-$t''$-$J$ model at $T=0$ \cite{tohy2}. However, results show
persistence of the pseudogap in the nodal direction even in the
overdoped regime. Also, the temperature behavior of the pseudogap
in spectral function (SF) has not been clarified yet.

In this Letter we present results of a numerical study of the
$t$-$t'$-$J$ model at finite $T>0$ which reveal that most qualitative
features of ARPES results are a nontrivial consequence of strong
correlations and do not require a doping-dependent model parameters,
in particular not the closing of the Mott-Hubbard gap. We show that
the pseudogap vanishes with doping in the nodal direction as well as above
the pseudogap temperature $T^*$ being closely tied to the disappearance of
longer-range AFM correlations. In the same way the FS transforms from a
pocket-like into a continuous large one. Further, our results allow for
the interpretation of two effective subbands both emerging from an AFM splitting
of a correlated (upper Hubbard) band. Numerically evaluated
$\sigma(\omega)$ can be also explained within the same pseudogap
scenario.

In the following we study the $t$-$t'$-$J$ model
\begin{equation}
H=-\sum_{i,j,s}t_{ij}\tilde{c}^\dagger_{js}\tilde{c}_{is}
+J\sum_{\langle ij\rangle}({\bf S}_i\cdot {\bf S}_j-\frac{1}{4}
n_in_j), \label{eq1} 
\end{equation}
where $\tilde{c}^\dagger_{is}$ are projected fermionic operators not
allowing the double occupancy of sites.  On a square lattice we
include besides the nearest-neighbor hopping $t_{ij}=t$ also the
second-neighbor hopping $t_{ij}=t^\prime$, where we take $t'=0.3 t$,
as relevant for EDC \cite{tohy1,tohy2}. We also fix $J=0.3~t$.

For the calculation of the SF, $A({\bf k},\omega)$, we use the
finite-temperature Lanczos method (FTLM) \cite{jprev}. It has been
shown that on small systems reachable by ED the calculation at $T>0$
gives more macroscopic-like results and gives a better controlled
approach to $T\to 0$ provided that $T>T_{fs}$ where $T_{fs}$ is the
finite-size temperature determined by the model and system size
\cite{jprev}.  For cases considered here with tilted square lattices of
$N=18,20$ sites and $N_e=0-4$ doped electrons we have $T_{fs} \sim
0.1~t$.  On small systems with fixed boundary conditions one can
consider only a discrete set of wavevectors ${\bf k}_l, l=1,N$. To
scan the whole Brillouin zone of ${\bf k}$ we employ twisted boundary
conditions by introducing hopping elements $t_{ij} \to \tilde t_{ij} =
t_{ij} ~\mathrm {exp}(i\vec{\theta}\cdot\vec{r}_{ij})$ in
Eq.~(\ref{eq1}), in this way reaching arbitrary momenta ${\bf k}={\bf
k}_l+\vec \theta$ \cite{tohy2}. Using FTLM we calculate the Green's
function $G({\bf k},\omega)$ and therefrom extract the self energy $\Sigma({\bf k},\omega)=\omega -\zeta_{\bf k}-\alpha/G({\bf k},\omega)$.
Borrowing the idea from cluster dynamical
mean-field approaches we smoothen $\Sigma({\bf k},\omega)$ over small
interval $\delta k \sim 0.3$ and then recalculate SF. For details we
refer to Ref.\cite{zeml2}.

\begin{figure}[htb]  
\centering 
\epsfig{file=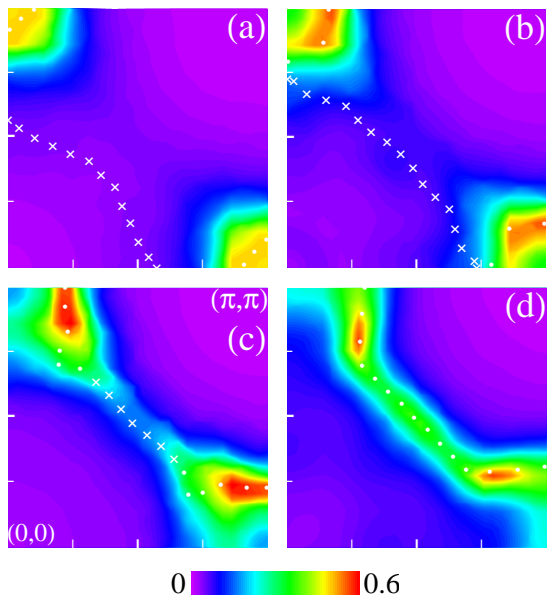,width=80mm,angle=0}
\caption{(Color online) Contour plot of $A({\bf k},\omega =0)$ representing the Fermi
surface for different electron doping: (a) $c_e=1/20$, (b) $c_e=2/20$,
(c) $c_e=3/20$, and (d) $c_e=4/18$. Results are shown for $T/t=0.1$. }
\label{fig1}   
\end{figure}

Let us start with the presentation of results for the FS evolution.
In Fig.~1 we display the scans of the SF at the chemical potential $A({\bf k},\omega =0)$, as
usual also in ARPES presentations of FS \cite{dama,armi2,mats,park}, in the
quarter of the first Brillouin zone for different electron dopings
$c_e=N_e/N$. Data in Fig.~1 are supplemented with
Fig.~2 where full $A({\bf k},\omega)$ are presented
along the AFM zone-boundary line.

At the lowest doping $c_e=1/20$ in Fig.~1a we get
quasiparticle weight forming a circle-like structure around the
$(\pi,0)$ point, well known as the electron pocket consistent with
several experimental and theoretical studies
\cite{armi2,kusu,tohy2,yuan}. The FS scan looks similar also for
$c_e=2/20$ in Fig.~1b except that the pocket gets more asymmetric
weight (opening from the $k_y=0$ side) around $(\pi,0)$. Within the
rest of the Brillouin zone, in particular along the AFM zone-boundary,
the SF shows a pseudogap at $\omega\sim 0$, as evident also in
Figs.~2a,b. In this doping regime the FS can be defined in a usual
way, i.e. as a ${\bf k}$-line of points where the quasiparticle peak in SF
crosses the chemical potential,
represented as white dots around $(\pi,0)$ in Figs.~1a,b. On the other hand,
white crosses in Figs.~1a-c indicate a ${\bf k}$-line where
the SF shows symmetric pseudogap. Note that the line of 
zeroes in the real part of the Green's function
determining the Luttinger sum rule is then represented by the
combination of both lines. It is indicative in Fig.~1b that already for $c_e=2/20$
both lines nearly touch as a precursor of a large FS formation.

As the doping is further increased, in Fig.~1c, the FS is formed almost
along the whole Luttinger line. The pockets are completely opened and
the transfer of the SF weight into the nodal region is observed,
although the pseudogap feature remains visible as seen in Fig.~2c. Finally,
at $c_e=4/18$ in Fig.~1d the pseudogap disappears also in the nodal
direction resulting in a large continuous FS. In addition, the SF in
Fig.~2d are well consistent with the usual quasiparticle peak crossing
the chemical potential. 

\begin{figure}[htb]  
\centering 
\epsfig{file=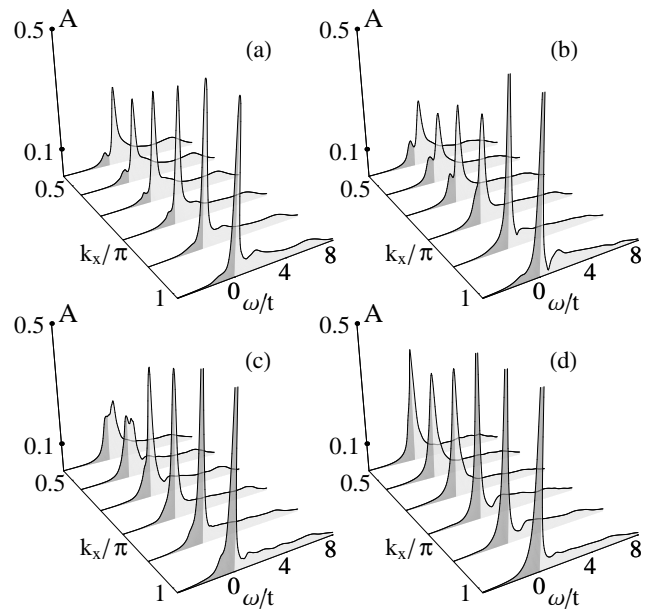,width=90mm,angle=0}
\caption{Spectral functions $A({\bf k},\omega)$ along the AFM zone-boundary
line. Dopings are the same as in Fig.~1 and $T/t=0.1$.}
\label{fig2}   
\end{figure} 

In Fig.~3 we focus in more detail on the doping and temperature
dependence of the pseudogap. We present
$A({\bf k},\omega)$ in the nodal point of the Luttinger surface
located as shown in the inset of
Fig.~3a. It is evident that a pronounced dip in SF at low doping
$c_e<0.15$ fills up with increasing doping and transforms into a usual
peak at $c_e=4/18$. The pseudogap is clearly manifested also in the
behavior of the corresponding self-energies $\Sigma''({\bf k},\omega)$
in Fig.~3b. Besides the overall linear dependence $|\Sigma''({\bf
k},\omega)| \propto \omega$ for $\omega>0$, characteristic for anomalous
(marginal) Fermi liquid and well established also in hole-doped
cuprates \cite{dama,zeml2}, additional narrow Lorentzian at $\omega
\sim 0$ reveals the pseudogap and its strength \cite{zeml2}. Again,
this feature vanishes at $c_e=4/18$, but as well disappears with
increasing $T$. The intensity (the area) $\Delta^2$ of the pseudogap contribution
in $\Sigma''({\bf k},\omega)$ is the measure of the pseudogap width,
i.e. $ \propto \Delta$ \cite{zeml2}. Therefore we present in Fig.~3c
$T$-dependence of $\Delta^2$ for different $c_e$. It is evident
that temperature $T^*$ above which the pseudogap
effectively disappears decreases with $c_e$ and exists only in
the underdoped regime with $c_e<4/18$. This clearly links pseudogap to
the presence
of longer-range AFM correlations in the system. In this sense, the pseudogap discussed
here is related to large (high-energy) pseudogap observed in
hole-doped cuprates \cite{dama}.

\begin{figure}[htb]  
\centering 
\epsfig{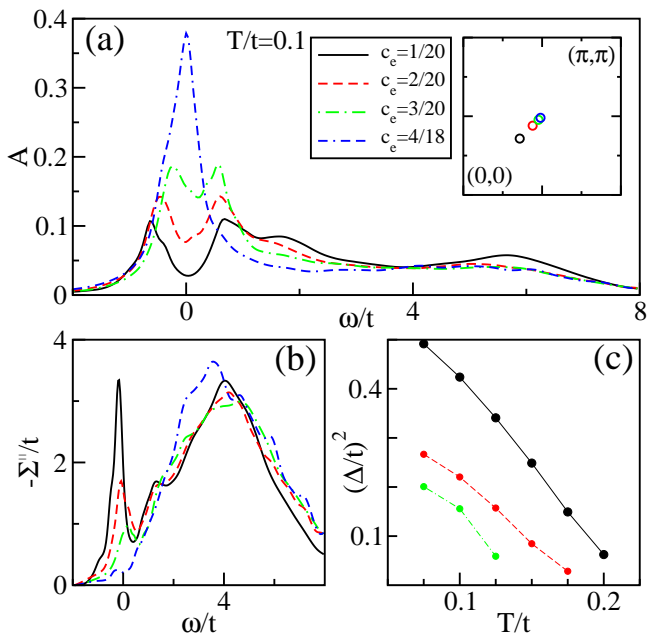}
\caption{(Color online) a) $A({\bf k},\omega)$ in the nodal point of the Luttinger
surface presented in the inset, b) corresponding self-energies
$\Sigma''({\bf k},\omega)$, and c) pseudogap intensity $\Delta^2$
vs. $T$ for different $c_e$.}
\label{fig3}   
\end{figure} 

In Fig.~4 we present a weight map of $A({\bf k},\omega )$ along
symmetry lines in the first Brillouin zone. Again, one can notice
strong electron-pocket contribution close to $(\pi,0)$ and evident
pseudogap in the nodal direction at lowest $c_e=1/20$ which becomes
less pronounced at higher doping and disappears at $c_e=4/18$. In
addition, it is interesting to follow effective bands which are clearly
seen in Fig.~4. At lowest $c_e=1/20$ one can well resolve
the two-band structure, at least along certain ${\bf k}$ directions,
e.g., along $(0,0)-(\pi/2,\pi/2)$, $(0,0)-(\pi,0)$ as well as
$(\pi,0)-(\pi/2,\pi/2)$. This feature dissolves with doping and
finally at $c_e=4/18$ transforms into a single band. Such a two-band
structure is very close to the one observed and analyzed in ARPES on EDC
\cite{mats,park}. The weaker band can be interpreted as the shadow-one
due to AFM order and its intensity can be linked to AFM order
parameter $\bar s$. In fact, results in Fig.~4 can be well
represented by effective bands of the form,
\begin{equation}
\epsilon_{\pm}({\bf k})=-4\tilde t' \gamma'_{\bf k} \pm \sqrt{(4 \tilde
t \gamma_{\bf k})^2 + w\bar s^2}, \label{eqe}
\end{equation}
where $\gamma_{\bf k}=(\cos k_x+\cos k_y)/2,~\gamma'_{\bf k}=
\cos k_x \cos k_y$ are
corresponding to first and second-nearest neighbor effective hopping
$\tilde t,\tilde t' $, respectively. The dispersion of the form
(\ref{eqe}) can be considered as an effective tight-binding band split
due to long-range AFM $\bar s \neq 0$ and can be analytically derived
from the $t$-$t'$-$J$ (or Hubbard) model using the memory-function approach
\cite{prel} whereby $\tilde t,\tilde t'$ are
renormalized by strong correlations. Very similar bands have been
recently used to describe observed ARPES spectra in SCCO \cite{park}.
 
In Fig.~4 we also plot fits of $\epsilon_{\pm}({\bf k})$, with
corresponding intensity of subbands shown in the scale from black (no
intensity) to white. We note that the main doping dependence emerges
through $\bar s=0.35-0$ which is the largest for low doping and zero at
$c_e=4/18$ while other parameters are only weakly doping dependent,
e.g. $\tilde t/t=0.3-0.6, -\tilde t'/t=0.1-0.2$ and $w=t^2$.

\begin{figure}[htb]  
\centering 
\epsfig{file=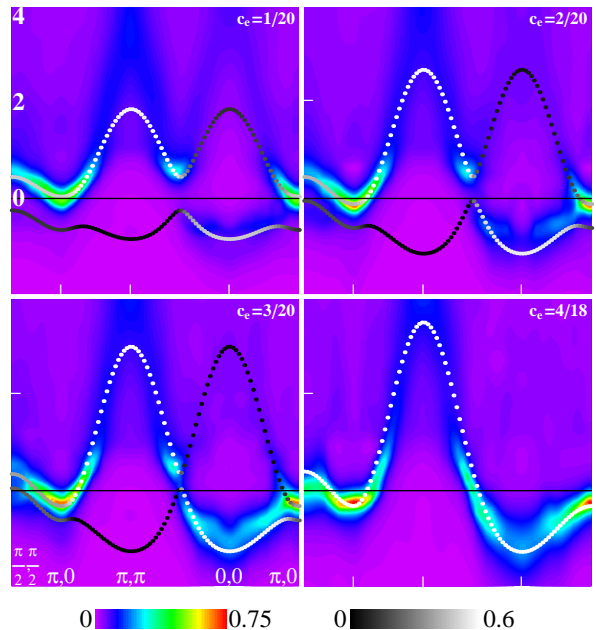,width=85mm,angle=0} 
\caption{(Color online) Weight map $A({\bf k},\omega)$ along symmetry lines in
the first Brillouin zone for different dopings and $T/t=0.1$. The fit to split subbands is shown
with the corresponding intensities.}
\label{fig4}   
\end{figure} 

Here we comment also on the asymmetry of high-energy dispersion ($\omega<0$)
between hole and electron-doped systems. From Fig.~4
we can clearly see that the
dispersion from $(\pi/2,\pi/2)$ to $(0,0)$ is a rather
standard one without significant kinks, but for $\omega>0$, which
corresponds to $\omega<0$ case of the hole-doped system in $t-t'-J$
models, a kink
behavior is seen along $(\pi/2,\pi/2)$ to $(\pi,\pi)$ as well as from 
$(\pi,0)$ to $(\pi,\pi)$. This difference is consistent with
related asymmetry observed in ARPES recently\cite{pan}.
The details will be given elsewhere\cite{zeml3}.

Finally, let us discuss the relation with the optical conductivity
$\sigma(\omega)$ in EDC \cite{park,onos} which also shows pronounced
pseudogap at low doping. First, we evaluate $\sigma(\omega)$ directly
by calculating the dynamical
current-current correlations in the ground state at $T=0$ and via FTLM
at $T>0$ \cite{jprev,zeml1}. Results for $T=0$ and $T/t =0.15~t$ at
different $c_e$ are presented in Fig.5. On the other hand, to
establish the relation with the pseudogap observed in SF we also
consider the simplest (decoupling) approximation in terms of SF neglecting
the vertex corrections,
\begin{equation}
\sigma_d(\omega)=\frac{2\pi}{\omega N} \sum_{\bf k} (v_{\bf k})^2
\int {d\omega^\prime} g(\omega,\omega')
A({\bf k},\omega^\prime) A({\bf k},\omega^\prime-\omega),
\label{eqsig}
\end{equation}
where $g(\omega,\omega')=f(\omega^\prime-\omega)-f(\omega^\prime)$ and
${\bf v}_{\bf k}$ are bare band velocities as determined by $t,t'$.
Numerically evaluated SF as presented
above are inserted into Eq.~(\ref{eqsig}) and results are plotted in Fig.~5 as black lines. A
nontrivial message following from Fig.~5 is that $\sigma_d(\omega)$
yields very good overall qualitative
and even quantitative agreement with the exactly evaluated
$\sigma(\omega)$ (at $T>0$), whereby features are plausibly broader in
$\sigma_d(\omega)$.

\begin{figure}[htb]  
\centering 
\epsfig{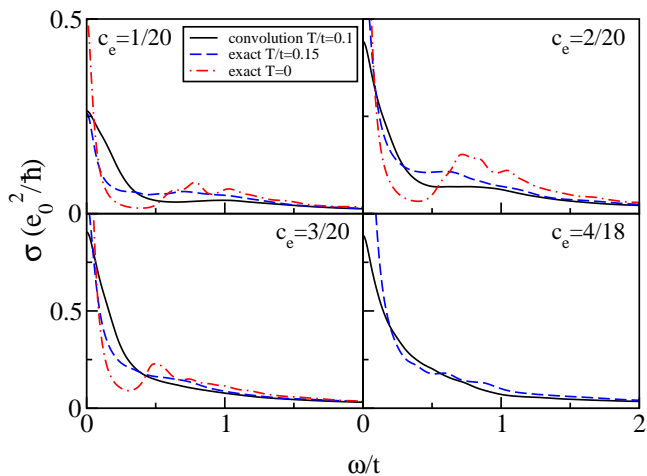}
\caption{(Color online) Optical conductivity $\sigma(\omega)$ as obtained from the 
small-system calculation at $T=0$ (red line), shown for
$c_e=1/20, 2/20, 3/20$, and $T/t=0.15$ (blue line). Black lines
represent $\sigma_d(\omega)$ calculated from the convolution of
spectral functions at $T/t=0.1$.}
\label{fig5}   
\end{figure} 

At low doping, $c_h=1/20, 2/20$ we can observe at $T/t=0.15~t$ a hump
in $\sigma(\omega)$ at $\omega /t\sim 0.7$, even more pronounced at
$T=0$, which can be interpreted as a manifestation of the pseudogap.
The same peak-dip feature is also resolved (although with somewhat
shifted peak) in convoluted $\sigma_d(\omega)$ which is the indication
that the pseudogap in conductivity has the same origin as the one in the SF. Furthermore,
$v_{\bf k}$ is the largest along the nodal part of the FS so it is
plausible that $\sigma(\omega)$ also tests the nodal
pseudogap. Further doping suppresses the peak-dip feature in Fig.~5 at
$c_e =4/18$ which can be related to the simultaneous pseudogap
disappearance in $A({\bf k},\omega)$. 

In conclusion, presented results for spectral functions and optical
conductivity show overall good qualitative and even quantitative
agreement with experimental results on EDC with different doping, both
for ARPES on NCCO and SCCO \cite{armi2,mats,park}, and
$\sigma(\omega)$ \cite{park,onos}. Since we have used the simplest prototype
model, Eq.(\ref{eq1}), for strongly correlated electrons with fixed
parameters, observed nontrivial phenomena as the Fermi surface
reconstruction, pseudogap and split bands appear as a direct consequence of
strong correlations. So there is no need to invoke a doping-dependent
Mott-Hubbard gap.

In comparison with more investigated hole-doped cuprates, EDC seem to
have an advantage that anomalous SF as well as $\sigma(\omega)$ can be
explained in a more straightforward way by invoking the long-range AFM
order $\bar s$, persisting in EDC materials up to $c_e=0.15$. In
particular, the double effective bands as seen in Fig.~4 at lower
doping as well as in recent ARPES experiments
\cite{mats,park} can be interpreted with a band splitting proportional
to $\bar s$, Eq.(\ref{eqe}), which disappears in the overdoped regime.

The pseudogap which is pronounced in SF along the zone diagonal and in
$\sigma(\omega)$, shows strong doping and temperature dependence,
whereby we can establish the pseudogap temperature $T^*(c_e)$
consistent with the one deduced from the optical conductivity
$\sigma(\omega)$ \cite{onos} but so far not analyzed with ARPES.
Both doping and $T$-dependence clearly establish longer-range AFM 
correlations as the origin of the pseudogap in EDC.

This work was supported by the Slovenian Research Agency under grant
PI-0044. T.T. acknowledges supports from the Next Generation Super Computing Project of Nanoscience Program, CREST, and Grant-in-Aid for Scientific Research from MEXT, Japan.

\end{document}